\documentclass[aps,preprint,superscriptaddress,showpacs,floatfix]{revtex4}%
\usepackage{amsfonts}
\usepackage{amsmath}
\usepackage{amssymb}
\usepackage{graphicx}
\usepackage{graphicx}%
\begin{document}
\title{Time-delayed coincidence technique for subnatural-width spectroscopy as an
interference phenomenon}
\author{R. N. Shakhmuratov}
\affiliation{Kazan Physical-Technical Institute, Russian Academy of Sciences,
10/7 Sibirsky Trakt, Kazan 420029 Russia}
\affiliation{Kazan Federal University, 18 Kremlyovskaya Street, Kazan 420008 Russia}
\author{F. G. Vagizov}
\affiliation{Kazan Federal University, 18 Kremlyovskaya Street, Kazan 420008 Russia}
\date{{ \today}}

\begin{abstract}
Single photon, emitted in a transition between two states, has a frequency
distribution of intensity, which is given by Lorentzian if the transition is
only naturally broadened and the period of observation $T$ is long compared to
the lifetime $T_{1}$ of the excited state. However, when the observation time
$T$ is short or comparable to $T_{1}$, the frequency spectrum is appreciably
broadened. If only the delayed part of the emitted radiation field is
detected, then the radiation spectrum does not change. However, if the
radiation field is transmitted through a resonant absorber and then detected,
the absorption/transmission spectrum of the delayed radiation field is narrowed.
We show that this narrowing is due to the interference of the spectral
components of the field. Experimental spectra of absorption of M\"{o}ssbauer
radiation, obtained by coincidence technique, confirm this conclusion.

\end{abstract}
\maketitle

\section{Introduction}

The intensity distribution of a line emitted in a transition between two
quantum states is given by Lorentzian if the line is only naturally broadened
and the period of observation $T$ is long. However, when time $T$ is short or
comparable to the lifetime $T_{1}$ of the excited state, the frequency
spectrum of the emitted radiation is appreciably broadened. Experiments with
gamma photons clearly demonstrated the spectrum broadening when $T<T_{1}$
\cite{Wu,Craig}. These experiments make use of the coincidence-M\"{o}ssbauer
technique, which could be summarized for $^{57}$Fe as follows. The source
nucleus, $^{57}$Co, undergoes electron capture to form a second excited state
of $^{57}$Fe whose lifetime is 12 ns. This state decays via two-photon cascade
emitting sequentially 122-keV and 14.4-kev photons. Detection of the 122-keV
photon signals the occupation of the 14.4-keV excited state of $^{57}$Fe whose
lifetime is 141 ns. This allows one to use the M\"{o}ssbauer spectroscopy
selecting only those 14.4-keV photons that are emitted during some preset time
interval after detecting 122-keV photon. Such a "time filtering" results in
substantial modification of the experimentally-observed
absorption/transmission-line shapes.

The coincidence-M\"{o}ssbauer spectroscopy was applied by Hamill and Hoy
\cite{Hoy} to improve the spectral resolution by taking spectra counting only
those gamma photons coming from the source nucleus that have lived longer than
one lifetime. Appreciable line narrowing was explained by the argument that
photons, emitted by nuclei that have lived longer time, have a better defined
energy. Similar technique was also applied in optical domain and named
time-biased coherent spectroscopy since it is based on selective discarding
that part of the signal arriving shortly after the excitation of the state to
be studied (for a review see \cite{Walther}). Significant linewidth reduction
in this technique have resulted in substantial improvements in precision. The
gain is attained when one observes the radiation from naturally decaying
states a certain time interval after they are populated and, therefore, limits
data collection to the set of atoms which has survived in the excited state
for a longer time than average. In optical domain, time-biased technique was
successfully applied in double resonance \cite{Ma68} and level-crossing
\cite{Series68,Walther74} experiments.

In this paper we show that spectral resolution in time-biased coherent
spectroscopy originates from the interference phenomena. While, in optical
domain, this is obvious conclusion since the line narrowing effects occur only
when the phase of the decaying signal is preserved by a measuring process
(see, for example, \cite{Metcalf}), in gamma domain the line narrowing is
usually explained by a decrease of the energy uncertainty of photons emitted
at later time \cite{Hoy}. In this paper we show that the spectrum of photons,
detected with appreciable delay after the excited state population, does not
experience narrowing. We also derive a simple formula for estimation of the
spectral resolution in optical and gamma domains, since the line narrowing is
described by similar expressions in both domains.

\section{Spectrum of emitted photon in time-biased measurements}

Lynch \textit{et al}. \cite{Lynch} empirically introduced an expression for
the source photon within a classical theory of gamma photon propagation in a
dense resonant absorber. A single-photon radiation field is presented as a
damped electric field%
\begin{equation}
a(t)=\Theta(t)e^{-i\omega_{S}t-\Gamma_{S}t/2}, \label{Eq1}%
\end{equation}
where the distance from the source is neglected, the amplitude of the field at
the input of the absorber is normalized to unity, time $t_{0}$ when the source
nucleus is formed in the excited state is defined as $t_{0}=0$, $\Theta(t)$ is
the Heaviside step function, $\omega_{S}$ is the frequency of the resonant
transition from the excited to the ground state, and $\Gamma_{S}$ is the
reciprocal of the mean life of the excited state $T_{1}$, i.e., $\Gamma
_{S}=1/T_{1}$.

If we adopt the Fourier transform of the form%
\begin{equation}
F(\omega)=%
{\displaystyle\int\limits_{-\infty}^{+\infty}}
f(t)e^{i\omega t}dt, \label{Eq2}%
\end{equation}
the Fourier transform of the radiation field amplitude (\ref{Eq1}) is%
\begin{equation}
A(\omega)=\frac{1}{\Gamma_{S}/2-i(\omega-\omega_{S})}, \label{Eq3}%
\end{equation}
and its intensity distribution $I(\omega)=A(\omega)A^{\ast}(\omega)$ is given
by Lorentzian function%
\begin{equation}
I(\omega)=\frac{1}{(\Gamma_{S}/2)^{2}+(\omega-\omega_{S})^{2}}. \label{Eq4}%
\end{equation}
Assume a shutter is available in gamma domain and we place this shutter
between the source and detector. In this gedankenexperiment we also suppose
that our detector is capable to measure the spectral content of the radiation
field. Below we consider two cases, i.e., (i) the shutter is open in a time
interval $(0,T)$ allowing detection only a leading edge of a single-photon wave
packet, and (ii) the shutter is open $(T,+\infty)$ to detect only the tail of
a single-photon wave packet. \begin{figure}[ptb]
\resizebox{0.8\textwidth}{!}{\includegraphics{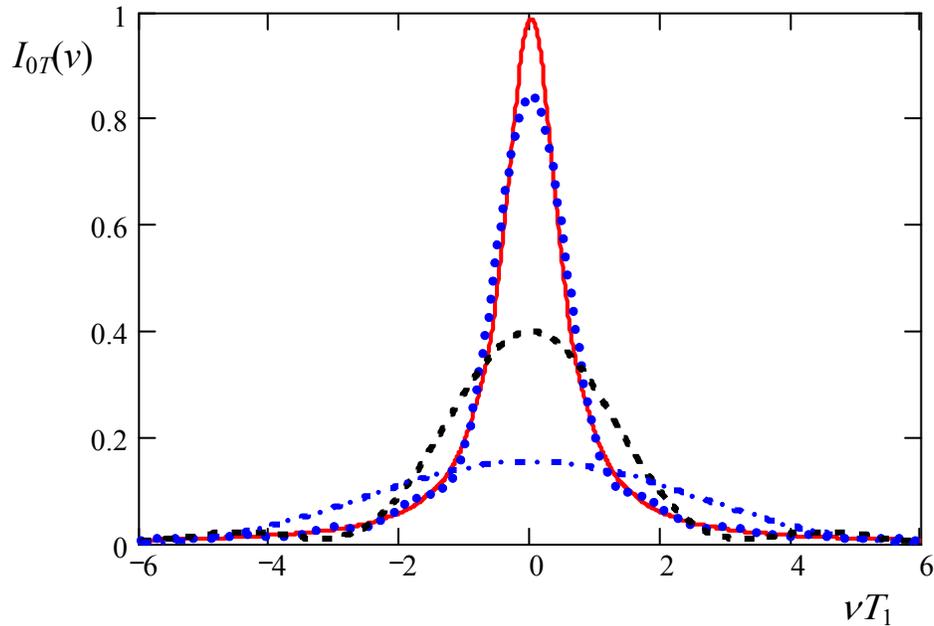}}\caption{ Power
spectrum of the source radiation field, seen by the frequency selective
detector in a thought experiment. The opening time $T$ of the shutter is
$10T_{1}$ (solid line in red), $5T_{1}$ (dotted line in blue), $2T_{1}$
(dashed line in black), and $T_{1}$ (dash-dotted line in blue).}%
\label{fig:1}%
\end{figure}

In the first case, the probability amplitude of the photon, transmitted by the
shutter, is%
\begin{equation}
a_{0T}(t)=a(t)[1-\Theta(t-T)], \label{Eq5}%
\end{equation}
and its Fourier transform%
\begin{equation}
A_{0T}(\omega)=\frac{1-e^{i(\omega-\omega_{S})T-\Gamma_{S}T/2}}{\Gamma
/2-i(\omega-\omega_{S})} \label{Eq6}%
\end{equation}
gives the following expression for the radiation intensity \cite{Wu}%
\begin{equation}
I_{0T}(\nu)=\frac{1+e^{-\Gamma_{S}T}-2e^{-\Gamma_{S}T/2}\cos\nu T}{(\Gamma
_{S}/2)^{2}+\nu^{2}}, \label{Eq7}%
\end{equation}
where $\nu=\omega-\omega_{S}$. For a long time interval when the shutter is
open, the radiation spectrum is close to Lorentzian (\ref{Eq4}), see Fig. 1,
solid line (in red), which corresponds to $T=10T_{1}$. With time-interval
shortening, the spectrum broadens. For example, for $T=T_{1}$ the spectrum
width increases $5.6$ times (see Fig.1, dash-dotted line, in blue). The drop
of intensity at the line center $\nu=0$ is due to discarding appreciable part
of a single-photon pulse and described by equation%
\begin{equation}
I_{0T}(0)=\frac{\left(  1-e^{-\Gamma_{S}T/2}\right)  ^{2}}{(\Gamma/2)^{2}}.
\label{Eq8}%
\end{equation}
For $T$ close to $T_{1}$, the intensity drops slightly more than two times
with respect to the line center when $\nu T=\pm\pi$. For this frequencies Eq.
(\ref{Eq7}) is simplified as follows%
\begin{equation}
I_{0T}(\pm\pi/T)=\frac{\left(  1+e^{-\Gamma_{S} T/2}\right)  ^{2}}{(\Gamma
_{S}/2)^{2}+(\pi/T)^{2}}. \label{Eq9}%
\end{equation}
Thus, when $T=T_{1}$ halfwidth at half-maximum of the radiation field
increases almost $\pi$ times.

If the shutter is open at time $T$ for a long time interval $(T,+\infty)$, the
probability amplitude of the photon transmitted by the shutter is%
\begin{equation}
a_{T\infty}(t)=\Theta(t-T)a(t). \label{Eq10}%
\end{equation}
Its Fourier transform%
\begin{equation}
A_{T\infty}(\nu)=\frac{e^{i\nu T-\Gamma_{S}T/2}}{\Gamma_{S}/2-i\nu}
\label{Eq11}%
\end{equation}
gives the following expression for the radiation intensity%
\begin{equation}
I_{T\infty}(\nu)=\frac{e^{-\Gamma T}}{(\Gamma_{S}/2)^{2}+\nu^{2}},
\label{Eq12}%
\end{equation}
which differs from Lorentzian profile (\ref{Eq4}) only by the numerical factor
$\exp(-T)$. Thus, there is no spectrum narrowing of photons, which are
detected at later time.

In gamma domain the shutters and spectrum selective detectors are not yet
available. Instead, time-delayed coincidence counting technique and resonant
absorbers are used \cite{Wu,Craig,Hoy}. While, absorption/transmission
spectrum broadening, detected by this technique, confirms the spectrum
broadening of a photon for the observation time interval $(0,T)$, the line
narrowing in the time interval $(T,+\infty)$ could be explained only by the
interference phenomenon since the photon spectrum is not narrowed in delayed
detection. This point will be discussed in Sec. IV. Below, we summarize the
results of time-biased coherent spectroscopy in optical domain showing that
interference plays a crucial role to improve resolution in level-crossing experiments.

\section{Level-crossing experiments}

In level-crossing experiments, atomic vapor in a cell is illuminated by a
short pulse, which is produced by a vapor lamp chopped by a Kerr cell or by a
pulsed laser. If two closely spaced excited states $\left\vert e_{1}%
\right\rangle $ and $\left\vert e_{2}\right\rangle $ of atoms are populated in
a coherent mixture, the atom evolution is described by the wave function%
\begin{equation}
\left\vert \Psi(t)\right\rangle =C_{1}(t)\left\vert e_{1}\right\rangle
+C_{2}(t)\left\vert e_{2}\right\rangle +C_{0}\left\vert g\right\rangle ,
\label{Eq13}%
\end{equation}
where for a weak excitation we have $C_{0}\approx1$, $C_{n}(t)\approx\beta
_{n}e^{-i\omega_{n}t-\Gamma t/2}$, $n=1,2$, $\omega_{n}=E_{n}/\hbar$, $E_{n}$
is the energy of the $n$-th excited state, $\Gamma/2$ is the decay rate of the
probability amplitudes of the excited states, and $\beta_{n}$ is proportional
to the transition matrix element and amplitude of the exciting field.
Intensity of fluorescence due to a decay of the excited states $\left\vert
e_{1}\right\rangle $ and $\left\vert e_{2}\right\rangle $ is proportional to
$\left\vert V\left[  C_{1}(t)+C_{2}(t)\right]  \right\vert ^{2}$, where $V$ is
the dipole interaction with vacuum modes. Therefore, the field intensity
emited in a $4\pi$ angle is%
\begin{equation}
I(t)=2\left\vert V\right\vert ^{2}(1+\alpha\cos\Omega t)e^{-\Gamma t},
\label{Eq14}%
\end{equation}
where $\hbar\Omega=E_{2}-E_{1}$ is the energy difference of the excited states
$\left\vert e_{1}\right\rangle $ and $\left\vert e_{2}\right\rangle $,
$\alpha$ equals $+1$ or $-1$ depending on the transition matrix elements,
geometry of excitation and observation \cite{Series68,Walther74,Metcalf}.
Below, for simplicity we take $\alpha=1$. Immediately after the excitation the
radiation fields emitted by the excited states $\left\vert e_{1}\right\rangle
$ and $\left\vert e_{2}\right\rangle $ are in phase and they interfere
constructively for $\alpha=+1$. When $\omega t=\pi$ the interference becomes
destructive. Thus, the emitted field reveals time bitting with a period
$2\pi/\omega$. However, because it was difficult to measure bitting directly,
time integrated signals were collected in level mixing experiments.

A number of photons, detected, for example, at the right angle to the
excitation direction during a long time period after the radiation pulse, has
a simple dependence on the splitting $\hbar\Omega$ (see also Fig. 2)%
\begin{equation}
n_{0\infty}(\Omega)=%
{\displaystyle\int\limits_{0}^{+\infty}}
I(t)dt=N_{0}\left(  1+\frac{\Gamma^{2}}{\Omega^{2}+\Gamma^{2}}\right)  ,
\label{Eq15}%
\end{equation}
where $N_{0}=2\left\vert V\right\vert ^{2}/\Gamma$. The interference of the
fields emitted by two excited states results in the second term in the
brackets in Eq. (\ref{Eq15}). Since the coherence of the fields is limited by
spectral width of each spectral component, which is $\Gamma$, the contribution
of the interference term drops when the distance between them, $\Omega$
becomes larger than $\Gamma$. This width limits the spectral resolution of the
position of level crossing.
\begin{figure}[ptb]
\resizebox{0.8\textwidth}{!}{\includegraphics{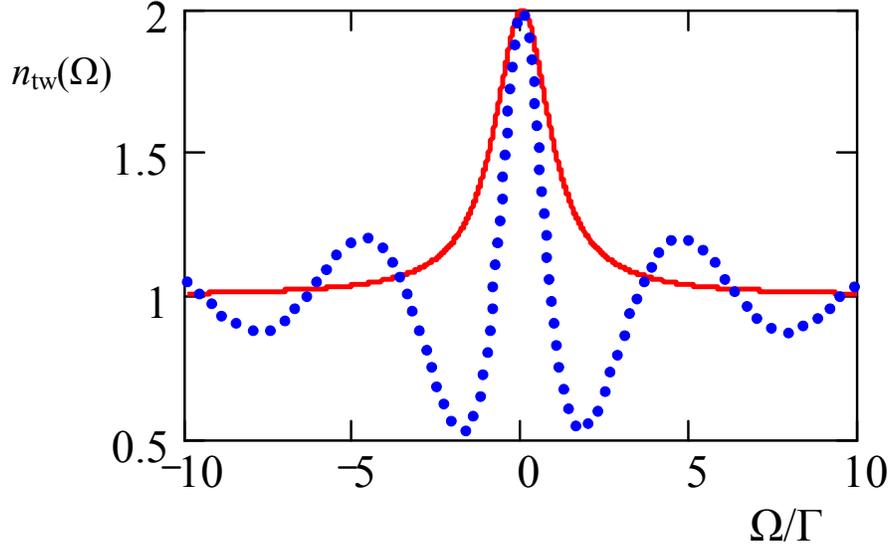}}\caption{Dependence of
the number of photon counts $n_{\text{tw}}(\Omega)$ on the excited states
splitting $\hbar\Omega$. Solid line (in red) shows the case when counts are
collected from time $0$ immediately after the excitation up to infinite time.
Dotted line (in blue) shows the case when data are collected from time
$t=T_{1}=1/\Gamma$ to infinity. Number of counts are normalized to $N_{0}$ in
both plots and exponential factor in Eq. (\ref{Eq4}) describing dotted line
(in blue) is disregarded in the plot for visualization. Index tw in
$n_{\text{tw}}(\Omega)$ means time window.}%
\label{fig:2}%
\end{figure}

If we discard photon counts in a time window $(0,T)$ and collect data only in
the interval $(T,+\infty)$, then we have%
\begin{equation}
n_{T\infty}(\Omega)=N_{0}e^{-\Gamma T}\left(  1+\Gamma\frac{\Gamma\cos\Omega
T-\Omega\sin\Omega T}{\Omega^{2}+\Gamma^{2}}\right)  . \label{Eq16}%
\end{equation}
The plot of $n_{T\infty}(\Omega)$ for $T=T_{1}=1/\Gamma$ is shown in Fig. 2 by
dotted line in blue. While without data discarding halfwidth of the line in
Fig. 2 is $\Gamma$, with data discarding the line narrows. To estimate line
narrowing for any delay time $T$, we express Eq. (\ref{Eq16}) as%
\begin{equation}
n_{T\infty}(\Omega)=N_{0}e^{-\Gamma T}\left[  1+\frac{\Gamma\cos(\Omega
T+\phi)}{\sqrt{\Omega^{2}+\Gamma^{2}}}\right]  , \label{Eq17}%
\end{equation}
where $\phi=\tan^{-1}\left(  \Omega/\Gamma\right)  $. When $\Omega=0$, the
number of photon counts has maximum value $n_{T\infty}(0)=2N_{0}\exp(-\Gamma
T)$. First minima appear when $\phi\pm\Omega T=\pm\pi$ (see Fig. 2). For
$T>2T_{1}$, this condition is approximately satisfied when%
\begin{equation}
\Omega_{\min}=\pm\frac{\pi\Gamma}{1+\Gamma T}. \label{Eq18}%
\end{equation}
Half width of the bell-shape central maximum of the dotted curve
in Fig. 2, which is described by Eq. (\ref{Eq17}), can be estimated as a
halfway between module of $\Omega_{\min}$ and $\Omega=0$, which is
\begin{equation}
\gamma_{\mathrm{coh}}=\frac{\pi/2}{T_{1}+T}. \label{Eq19}%
\end{equation}
If $T=2T_{1}$, this width, which is $\pi/6T_{1}$, reduces 2 times with respect
to the half width $1/T_{1}$ of the line measured without discarding data
(shown by solid line, in red, in Fig. 2).

We have to mention that approximation (\ref{Eq19}) is valid even for $T=T_{1}$
giving slightly underestimated halfwidth $0.78\Gamma$, which is by 11\%
smaller than actual width $0.87\Gamma$. Line narrowing originates from
interference of the fields emitted by two closely-spaced excited states if we
fix time $T$ when we start data collection. The longer this time, the narrower
the line is. For example, in experiments with excitation by short laser pulses
\cite{Walther74}, the fluorescent light was observed in time intervals which
were delayed up to seven lifetimes after excitation. The minimum linewidth
observed in these experiments was 6 times smaller than the natural width. Line
narrows, if one allows to acquire appreciable phase difference $\Omega T$ of
the excited states before data collection. \begin{figure}[ptb]
\resizebox{0.5\textwidth}{!}{\includegraphics{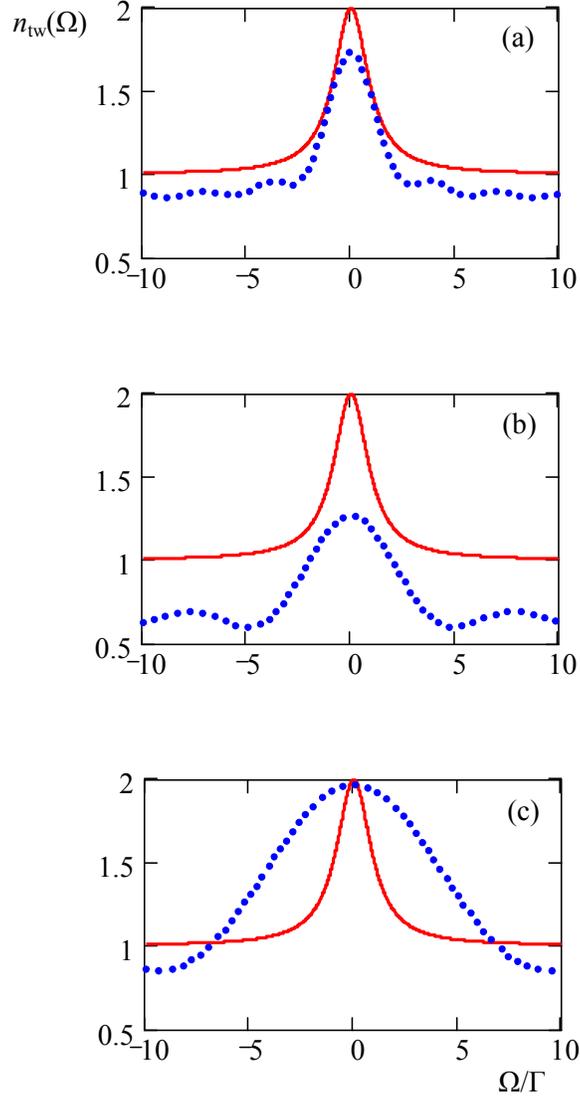}}\caption{Dependence of
the number of photon counts on the splitting $\hbar\Omega$. Solid line (in
red) shows the case when counts are collected in the infinite time interval
starting immediately after the excitation at $t=0$. Dotted line (in blue)
shows the case when data are collected from time $t=0$ to $t=T$. Number of
counts are normalized to $N_{0}$. Time $T$ is equal to $2T_{1}$ in (a),
$T_{1}$ in (b), and $T_{1}/2$ in (c). The values of $n_{0T}(\Omega)$ in (c)
(dotted line in blue) are magnified by a factor $2.5$ for visualization.}%
\label{fig:3}%
\end{figure}

If data collection would start at $t_{\mathrm{start}}=0$, i.e., just immediately
after the excitation, and terminate at $t_{\mathrm{stop}}=T$, the number of
counts is described by%
\begin{equation}
n_{0T}(\Omega)=n_{0\infty}(\Omega)-n_{T\infty}(\Omega). \label{Eq20}%
\end{equation}
The plots of $n_{0T}(\Omega)$ for different values of $T$ and their comparison
with $n_{0\infty}(\Omega)$ are shown in Fig. 3. For long time interval, for
example, $T=2T_{1}$ as in Fig. 3(a), the line almost coincides with that,
which is collected infinite time. With shortening of data-collection time
interval, the line broadens.

To analyse the line broadening with shortening of $T$, we express Eq
(\ref{Eq20}) as%
\begin{equation}
n_{0T}(\Omega)=\left(  1-e^{-\Gamma T}\right)  n_{0\infty}(\Omega
)+n_{\mathrm{broad}}(\Omega). \label{Eq21}%
\end{equation}
where%
\begin{equation}
n_{\mathrm{broad}}(\Omega)=N_{0}e^{-\Gamma T}\frac{\Gamma^{2}(1-\cos\Omega
T)+\Gamma\Omega\sin\Omega T}{\Omega^{2}+\Gamma^{2}}. \label{Eq22}%
\end{equation}
The plots of narrow $n_{\mathrm{narrow}}(\Omega)=\left(  1-e^{-\Gamma
T}\right)  n_{0\infty}(\Omega)$ and broad $n_{\mathrm{broad}}(\Omega)$ parts
of the spectrum together with their sum $n_{0T}(\Omega)$ are shown in Fig. 4.
The broad part contains a narrow hole at the center. It can be shown by simple
algebra, that the hole is fully compensated by the narrow part of the spectrum
because $n_{\mathrm{narrow}}(\Omega)$ is proportional to $\Gamma^{3}T/\left(
\Omega^{2}+\Gamma^{2}\right)  $, while $n_{\mathrm{broad}}(\Omega)$ is
proportional to $\Gamma\Omega^{2}T/\left(  \Omega^{2}+\Gamma^{2}\right)  $,
and their sum is equal to $N_{0}\Gamma T=N_{0}T/T_{1}$ when $\Omega
\rightarrow0$ and $T\ll T_{1}$. The wings of the broad line take minimum
values when $\Omega T=\pm3\pi/2$. Therefore, half width at halfmaximum of the
central line of the spectrum $n_{0T}(\Omega)$ equals $3\pi/4T$ when $T<2T_{1}$.

Broadening of the line $n_{0T}(\Omega)$ is caused by the spectrum broadening
$I_{0T}(\nu)$ of the truncated photon, see Sec. II, Eq. (\ref{Eq7}), since if
one measures photon counts in time window $0-T$ in level crossing experiments,
the spectrum of emitted photons is broadened. Below, we show the similarity
between $n_{0T}(\Omega)$ and $I_{0T}(\nu)$. Photon spectrum (\ref{Eq7}) can be
also expressed as a sum of narrow and broad components, i.e.,%
\begin{equation}
I_{0T}(\nu)=I_{\mathrm{narrow}}(\nu)+I_{\mathrm{broad}}(\nu), \label{Eq23}%
\end{equation}
where%
\begin{equation}
I_{\mathrm{narrow}}(\nu)=\frac{\left(  1-e^{-\Gamma_{S}T/2}\right)  ^{2}}%
{\nu^{2}+(\Gamma_{S}/2)^{2}}, \label{Eq24}%
\end{equation}%
\begin{equation}
I_{\mathrm{broad}}(\nu)=2e^{-\Gamma_{S}T/2}\frac{1-\cos\nu T}{\nu^{2}%
+(\Gamma_{S}/2)^{2}}. \label{Eq25}%
\end{equation}
\begin{figure}[ptb]
\resizebox{0.5\textwidth}{!}{\includegraphics{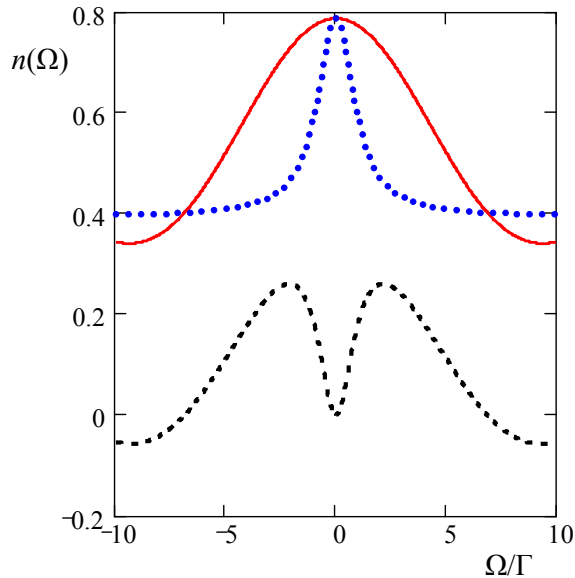}}\caption{(color on
line) Dependence of the number of photon counts on the splitting $\hbar\Omega
$. Solid line (in red) shows the case when counts are collected in time
interval $T=T_{1}/2$ starting immediately after the excitation at $t=0$.
Dotted line (in blue) is the narrow part of the spectrum $n_{\mathrm{narrow}%
}(\Omega)$ (see the text for details). Dashed line (in black) is the broad
part of the spectrum $n_{\mathrm{broad}}(\Omega)$.}%
\label{fig:4}%
\end{figure}Frequency dependencies of $I_{0T}(\nu)$, $I_{\mathrm{narrow}}%
(\nu)$, and $I_{\mathrm{broad}}(\nu)$ are very similar to those of
$n_{0T}(\Omega)$, $n_{\mathrm{narrow}}(\Omega)$, and $n_{\mathrm{broad}%
}(\Omega)$, respectively, shown in Fig. 4. First minima of intensity
$I_{0T}(\nu)$ are localized at $\nu=\pm2\pi/T$. Therefore, halfwidth at
halfmaximum of $I_{0T}(\nu)$ is $\pi/T$ for $T<1/\Gamma_{S}$. This result also
follows from simple properties of the Fourier transforms. Since, the
probability amplitude of the truncated photon, Eq. (\ref{Eq5}), is a product
of $a(t)$, Eq. (\ref{Eq1}), and a rectangular function $f_{\mathrm{R}%
}(t)=\Theta(t)[1-\Theta(t-T)]$, Fourier transform of Eq. (\ref{Eq5}) is a
convolution of the Fourier transforms of the probability amplitude of the
nontruncated photon $A(\omega)$ and rectangular function $F_{R}(\omega)$,
i.e.,%
\begin{equation}
A_{0T}(\nu)=\frac{1}{2\pi}%
{\displaystyle\int\limits_{-\infty}^{+\infty}}
A(\nu_{1})F_{R}(\nu-\nu_{1})d\nu_{1}, \label{Eq26}%
\end{equation}
where $\nu=\omega-\omega_{0}$ and%
\begin{equation}
F_{R}(\nu)=Te^{i\nu T/2}\frac{\sin\nu T/2}{\nu T/2}. \label{Eq27}%
\end{equation}
For $T\ll T_{1}$, halfwidth at halfmaximum of $F_{R}(\nu)$ is $\pi/T$, which
is much larger than that of $A(\nu_{1})$. Therefore, the width of $F_{R}(\nu)$
defines the width of truncated photon $A_{0T}(\nu)$ for short observation time
$T$.

\section{Coincidence-M\"{o}ssbauer spectroscopy}

Lynch \textit{et al}. \cite{Lynch} proposed to describe time dependence of the
radiation field amplitude of the source, Eq. (\ref{Eq1}), which is transmitted
through the absorber of physical thickness $L$, by inverse Fourier
transformation%
\begin{equation}
a_{\mathrm{out}}(t-t_{0})=\frac{1}{2\pi}%
{\displaystyle\int\limits_{-\infty}^{+\infty}}
A_{\mathrm{out}}(\omega)e^{-i\omega(t-t_{0})}d\omega, \label{Eq28}%
\end{equation}
where $A_{\mathrm{out}}(\omega)=A(\omega)\exp[-\alpha_{A}(\omega)L/2]$ is the
radiation spectrum at the exit of the absorber, $A(\omega)$ is the spectrum of
the incident radiation field [see Eq. (\ref{Eq3})], $t_{0}$ is time when the
excited state nucleus is formed in the source, and $\alpha(\omega)$ is the
complex dielectric constant, which describes frequency dependent absorption
and phase shift of the field in the sample. For the absorber with a single
resonance line, $\alpha_{A}(\omega)$ is%
\begin{equation}
\alpha_{A}(\omega)=\frac{i\Gamma_{A}T_{A}/2}{\omega-\omega_{A}+i\Gamma_{A}/2},
\label{Eq29}%
\end{equation}
where $\omega_{A}$ and $\Gamma_{A}$ are the resonant frequency and absorption
linewidth of nuclei, $T_{A}=n_{A}L\sigma f_{A}$ is optical thickness of the
absorber, $n_{A}$ is the number of nuclei per unit area of the absorber,
$\sigma$ is the resonance absorption cross section, and $f_{A}$ is the
recoilless fraction of gamma ray absorption. If we count a number of photons,
detected within long time intervals of the same duration for different values
of the detuning from resonance, $\Delta=\omega_{S}-\omega_{A}$, we obtain
M\"{o}ssbauer spectrum showing the dependence of transmitted radiation
intensity on $\Delta$. This number of counts is%
\begin{equation}
n_{\infty}(\Delta)=%
{\displaystyle\int\limits_{-\infty}^{+\infty}}
n_{\mathrm{out}}(t-t_{0})dt, \label{Eq30}%
\end{equation}
where $n_{\mathrm{out}}(t-t_{0})=$ $\left\vert a_{\mathrm{out}}(t-t_{0}%
)\right\vert ^{2}$. If we express $n_{\mathrm{out}}(t-t_{0})$ as%
\begin{equation}
n_{\mathrm{out}}(t-t_{0})=\frac{1}{(2\pi)^{2}}%
{\displaystyle\int\limits_{-\infty}^{+\infty}}
d\omega_{1}%
{\displaystyle\int\limits_{-\infty}^{+\infty}}
d\omega_{2}A_{\mathrm{out}}(\omega_{1})A_{\mathrm{out}}^{\ast}(\omega
_{2})e^{i(\omega_{2}-\omega_{1})(t-t_{0})}, \label{Eq31}%
\end{equation}
then, from Eq. (\ref{Eq30}) and definition of Dirac delta function%
\begin{equation}
\frac{1}{2\pi}%
{\displaystyle\int\limits_{-\infty}^{+\infty}}
e^{i(\omega_{2}-\omega_{1})(t-t_{0})}dt_{0}=\delta(\omega_{1}-\omega_{2}),
\label{Eq32}%
\end{equation}
we obtain%
\begin{equation}
n_{\infty}(\Delta)=\frac{1}{2\pi}%
{\displaystyle\int\limits_{-\infty}^{+\infty}}
d\omega A_{\mathrm{out}}(\omega)A_{\mathrm{out}}^{\ast}(\omega), \label{Eq33}%
\end{equation}
which is well known expression for transmitted radiation in M\"{o}ssbauer
spectroscopy \cite{Book}. Here, nonresonant absorption is disregarded. Recoil
processes in nuclear emission are not taken into account assuming that
recoilless fraction of the source emission (Debye-Waller factor) is $f_{S}=1$.
These processes can be easily taken into account in experimental data analysis.

\subsection{Photon counts in a short time interval starting immediately after
formation of the excited state nucleus in the source}

In time-delayed coincidence technique we are able to collect photon counts,
which are detected in a time interval $(t_{0},t_{0}+T)$. These counts satisfy
the equation
\begin{equation}
n_{0T}(\Delta)=%
{\displaystyle\int\limits_{t_{0}}^{t_{0}+T}}
n_{\mathrm{out}}(t-t_{0})dt. \label{Eq34}%
\end{equation}
Substituting integral representation of $a_{\mathrm{out}}(t-t_{0})$ and
$a_{\mathrm{out}}^{\ast}(t-t_{0})$ via Fourier transforms $A_{\mathrm{out}%
}(\omega)$ and $A_{\mathrm{out}}^{\ast}(\omega)$, Eq. (\ref{Eq28}), into
$n_{\mathrm{out}}(t-t_{0})=$ $a_{\mathrm{out}}(t-t_{0})a_{\mathrm{out}}^{\ast
}(t-t_{0})$ and taking into account the relation%
\begin{equation}%
{\displaystyle\int\limits_{t_{0}}^{t_{0}+T}}
e^{i(\omega_{2}-\omega_{1})(t-t_{0})}dt=Te^{i(\omega_{2}-\omega_{1}%
)T/2}\frac{\sin[(\omega_{2}-\omega_{1})T/2]}{(\omega_{2}-\omega_{1})T/2},
\label{Eq35}%
\end{equation}
we obtain%
\begin{equation}
n_{0T}(\Delta)=\frac{1}{(2\pi)^{2}}%
{\displaystyle\int\limits_{-\infty}^{+\infty}}
d\omega_{1}%
{\displaystyle\int\limits_{-\infty}^{+\infty}}
d\omega_{2}A_{\mathrm{out}}(\omega_{1})A_{\mathrm{out}}^{\ast}(\omega
_{2})F_{R}(\omega_{2}-\omega_{1}), \label{Eq36}%
\end{equation}
where $F_{R}(\omega_{2}-\omega_{1})$ is the function, which is defined in Eq.
(\ref{Eq27}) and coincides with that in the right hand side of Eq.
(\ref{Eq35}). Thus, in time-delayed coincidence technique we measure the
interference of spectral components $A_{\mathrm{out}}(\omega_{1})$ and
$A_{\mathrm{out}}^{\ast}(\omega_{2})$, which is governed by spectral function
$F_{R}(\omega_{2}-\omega_{1})$ whose width is defined by duration $T$ of the
observation time window.

Equation (\ref{Eq33}) is consistent with Parseval's theorem, which is written
as
\begin{equation}%
{\displaystyle\int\limits_{-\infty}^{+\infty}}
a_{\mathrm{out}}(t-t_{0})a_{\mathrm{ou}t}^{\ast}(t-t_{0})dt=\frac{1}{2\pi}%
{\displaystyle\int\limits_{-\infty}^{+\infty}}
A_{\mathrm{out}}(\omega)A_{\mathrm{out}}^{\ast}(\omega)d\omega. \label{Eq37}%
\end{equation}
However, if we limit time integration interval to $(0,T)$, then the relation
between $a_{\mathrm{out}}(t-t_{0})a_{\mathrm{ou}t}^{\ast}(t-t_{0})$ and
$A_{\mathrm{out}}(\omega)A_{\mathrm{out}}^{\ast}(\omega)$ changes to Eq.
(\ref{Eq36}), where we have interference of spectral components
$A_{\mathrm{out}}(\omega_{1})$ and $A_{\mathrm{out}}^{\ast}(\omega_{2})$
governed by the spectral function $F_{R}(\omega_{2}-\omega_{1})$. This change
can be explained by the spectral broadening of the radiation field at the exit
of the absorber if we shorten the observation time interval. According to Eq.
(\ref{Eq26}) the spectrum broadening is described by equation%
\begin{equation}
\mathcal{A}_{0T}(\nu)=\frac{1}{2\pi}%
{\displaystyle\int\limits_{-\infty}^{+\infty}}
A_{\mathrm{out}}(\nu_{1})F_{R}(\nu-\nu_{1})d\nu_{1}. \label{Eq38}%
\end{equation}
For the spectrum of the radiation field at the exit of the absorber, which is
calculated in a time interval $(0,T)$, Parseval's theorem holds, i.e.,%
\begin{equation}
n_{0T}(\Delta)=\frac{1}{2\pi}%
{\displaystyle\int\limits_{-\infty}^{+\infty}}
d\omega\mathcal{A}_{0T}(\omega)\mathcal{A}_{0T}^{\ast}(\omega). \label{Eq39}%
\end{equation}

To facilitate numerical calculation of the integrals in Eq. (\ref{Eq36}) we
present them with the help of substitution $\omega_{\pm}=\omega_{2}\pm
\omega_{1}$ as%
\begin{equation}
n_{0T}(\Delta)=\frac{1}{2(2\pi)^{2}}%
{\displaystyle\int\limits_{-\infty}^{+\infty}}
d\omega_{+}%
{\displaystyle\int\limits_{-\infty}^{+\infty}}
d\omega_{-}A_{\mathrm{out}}\left(  \frac{\omega_{+}-\omega_{-}}{2}\right)
A_{\mathrm{out}}^{\ast}\left(  \frac{\omega_{+}+\omega_{-}}{2}\right)
F_{R}(\omega_{-}). \label{Eq40}%
\end{equation}
It is possible to verify the accuracy of this expression if we calculate
numerically the integral in equation (\ref{Eq34}) for $n_{0T}(\Delta)$.
Analytical expression for the amplitudes $a_{\mathrm{out}}(t-t_{0})$ and
$a_{\mathrm{out}}^{\ast}(t-t_{0})$ in this integral can be found in Ref.
\cite{Lynch}, where they were calculated for $\Gamma_{A}=\Gamma_{S}$. For
practical use when $\Gamma_{A}\neq\Gamma_{S}$, the condition which is often
met in experiment, we also calculated analytically $a_{\mathrm{out}}(t-t_{0})$
for $\Gamma_{A}\neq\Gamma_{S}$. The result is%
\begin{equation}
a_{\mathrm{out}}(t)=\Theta(t)e^{-(\Gamma_{A}/2+i\omega_{A})t}%
{\displaystyle\sum\limits_{n=0}^{+\infty}}
\left[  s\left(  \frac{t}{b}\right)  ^{1/2}\right]  ^{n}J_{n}\left(
2\sqrt{bt}\right)  , \label{Eq41}%
\end{equation}
or%
\begin{equation}
a_{\mathrm{out}}(t)=\Theta(t)e^{-(\Gamma_{S}/2+i\omega_{S})t}\left[
e^{-b/s}-e^{-st}%
{\displaystyle\sum\limits_{n=1}^{+\infty}}
\left(  -b/s\right)  ^{n}\frac{J_{n}\left(  2\sqrt{bt}\right)  }{\left(
bt\right)  ^{n/2}}\right]  , \label{Eq42}%
\end{equation}
where $s=(\Gamma_{A}-\Gamma_{S})-i\Delta$, $b=T_{A}\Gamma_{A}/4$, and
$J_{n}(x)$ is the $n$-th order Bessel function, see also Ref. \cite{Varquaux}.

Moreover, it is possible to calculate the amplitude of the radiation field at
the exit of the absorber with the help of response function technique
\cite{Helisto}, which gives%
\begin{equation}
a_{\mathrm{out}}(t)=%
{\displaystyle\int\limits_{-\infty}^{+\infty}}
a(t-\tau)R(\tau)d\tau, \label{Eq43}%
\end{equation}
where $a(t)$ is defined in Eq. (\ref{Eq1}) and
\begin{equation}
R(t)=\delta(t)-e^{-(\Gamma_{A}+i\omega_{A})t}\Theta(t)\sqrt{\frac{b}{t}}%
J_{1}\left(  2\sqrt{bt}\right)  \label{Eq44}%
\end{equation}
is the response function (or Green function) of a single line absorber. Equation
(\ref{Eq43}) is reduced to%
\begin{equation}
a_{\mathrm{out}}(t)=a(t)\left[  e^{-st}J_{0}\left(  2\sqrt{bt}\right)  +s%
{\displaystyle\int\limits_{0}^{t}}
e^{-s\tau}J_{0}\left(  2\sqrt{b\tau}\right)  d\tau\right]  , \label{Eq45}%
\end{equation}
which is easy to calculate numerically because the intergrand is smooth and
not fast oscillating function.

The frequency dependencies of number of counts for infinite time interval
$n_{\infty}(\Delta)$, Eq. (\ref{Eq33}), and short time interval $n_{0T}%
(\Delta)$, Eq. (\ref{Eq36}), are shown in Fig. 5. Transmission spectrum
measured in a short time interval is appreciably broadened. The results of
spectrum calculation by formulae (\ref{Eq40}) and (\ref{Eq34}) are coincident.

\subsection{Analytical results for a short time interval}

To analyse analytically the spectrum broadening in coincidence-M\"{o}ssbauer
spectroscopy, similarly to the analysis presented in the previous section, we
consider absorber of moderate thickness $T_{A}$. Then, exponential function in
$A_{\mathrm{out}}(\omega)$ can be approximated as%
\begin{equation}
e^{-\alpha_{A}(\omega)L/2}=1-a_{1}x+a_{2}x^{2}+\varepsilon\left(  x\right)  ,
\label{Eq46}%
\end{equation}
where $x=$ $\alpha_{A}(\omega)L/2$, $a_{1}=0.9664$, $a_{2}=0.3536$, and
$\left\vert \varepsilon(x)\right\vert \leqslant3\cdot10^{-3}$ if $0\leqslant$
$\left\vert x\right\vert \leqslant\ln2$, see Ref. \cite{Abramowitz}. This
approximation is valid if $T_{A}\leqslant1.386$. \begin{figure}[ptb]
\resizebox{0.5\textwidth}{!}{\includegraphics{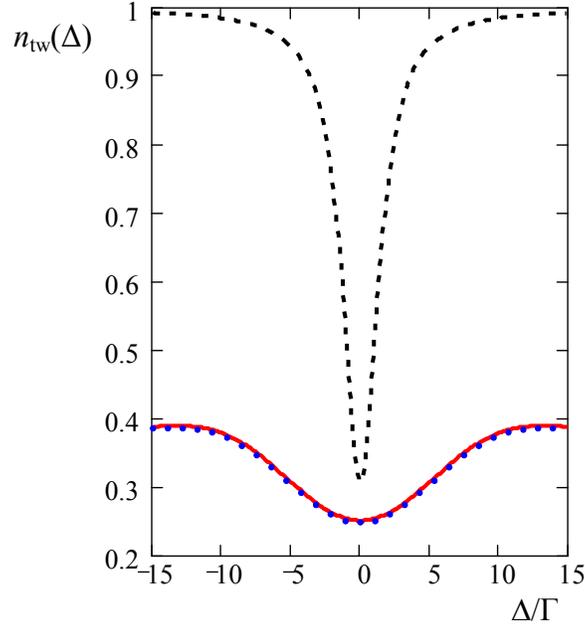}}\caption{Dependence of
number of counts on the detuning $\Delta$, which are collected in different
time windows $n_{\text{tw}}(\Delta)$. Here index tw means time window. Number
of counts are normalized to unity for $n_{\infty}(\pm\infty)$. Dashed line (in
black) shows the case when time window is $(0,\infty)$. Solid line (in red)
and dotted line (in blue) show the case when time widow $(0,T)$ is short,
i.e., $T=T_{1}/2=1/2\Gamma$. Spectra are plotted for $\Gamma_{A}=\Gamma
_{S}=\Gamma$ and optical thickness $T_{A}=4$. Solid line in red is plotted in
accord with Eq. (\ref{Eq34}), while dotted line in blue corresponds to Eq.
(\ref{Eq40}).}%
\label{fig:5}%
\end{figure}

Below, for simplicity we limit our consideration to the case $\Gamma
_{A}=\Gamma_{S}=\Gamma$. Then, with the help of approximation (\ref{Eq46}) we
calculated analytical expressions for the infinite-time spectrum $n_{\infty
}(\Delta)$, Eq. (\ref{Eq33}), and spectrum $n_{0T}(\Delta)$, which is detected
in a short time interval, Eq. (\ref{Eq36}). For the infinite-time spectrum the
result is%
\begin{equation}
n_{\infty}(\Delta)=\frac{1}{\Gamma}\left[  1-C_{1}\frac{\Gamma^{2}}{\Delta
^{2}+\Gamma^{2}}+C_{2}\frac{\Gamma^{4}}{\left(  \Delta^{2}+\Gamma^{2}\right)
^{2}}\right]  , \label{Eq47}%
\end{equation}
where $C_{1}=a_{1}T_{A}/2-a_{2}T_{A}^{2}/8$ and $C_{2}=a_{2}T_{A}^{2}/4$.
Comparison of the approximated result, Eq. (\ref{Eq47}), with exact dependence
$n_{\infty}(\Delta)$, Eq. (\ref{Eq33}), is shown in Fig. 6(a) for $T_{A}=1$.
The approximate dependence fits well the exact result even for absorbers with
moderate thickness.\begin{figure}[ptb]
\resizebox{0.5\textwidth}{!}{\includegraphics{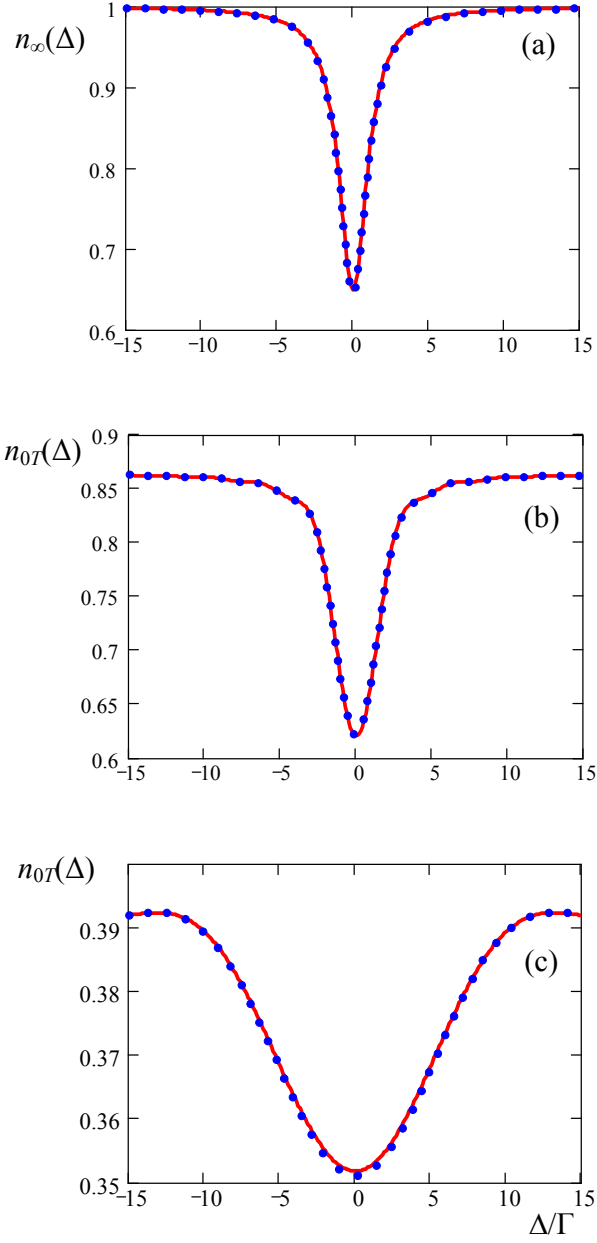}}\caption{(a)
Dependence of number of counts on the detuning $\Delta$, which are collected
in infinite time windows. Number of counts are normalized to unity for
$\Delta=\pm\infty$. Solid line (in red) shows the exact result, Eq.
(\ref{Eq33}), and dotted line (in blue) demonstrates the approximation Eq.
(\ref{Eq47}). Spectra $n_{0T}(\Delta)$ for $T=2T_{1}$ (b) and $T=T_{1}/2$ (c),
which are calculated with the help of exact equations (\ref{Eq34}),
(\ref{Eq41}), and(\ref{Eq42}) (solid line in red) and approximation Eq.
(\ref{Eq51}) (dotted line in blue). Optical thickness of the absorber is
$T_{A}=1$.}%
\label{fig:6}%
\end{figure}

It is interesting to notice that the spectrum $n_{\infty}(\Delta)$ consists of
two lines, one, with the coefficient $C_{1}$, has a halfwidth $\Gamma$, and
the other, with the coefficient $C_{2}$, has a halfwidth $\sqrt{\sqrt{2}%
-1}\Gamma=0.644\Gamma$, which is 1.55 times narrower. These lines are
subtracted (the narrow from the broad) and the result is the line with
increased width.

A halfwidth $\Gamma_{Th}$ of this broadened line $n_{\infty}(\Delta)$ can be
calculated from the equation $1-\Gamma n_{\infty}(\Gamma_{Th})=[1-\Gamma
n_{\infty}(0))]/2$, whose solution is
\begin{equation}
\Gamma_{Th}/\Gamma=\sqrt{\frac{A}{B}+\sqrt{1+\frac{A^{2}}{B^{2}}}},
\label{Eq48}%
\end{equation}
where $A=T_{A}a_{2}/2$ and $B=a_{1}-3T_{A}a_{2}/4$. For $T_{A}\leq1$ this
solution can be approximated as
\begin{equation}
\Gamma_{Th}/\Gamma=\sqrt{1+\frac{1}{2}(D+D^{2}+D^{3}+D^{4})}, \label{Eq49}%
\end{equation}
where $D=T_{A}a_{2}/a_{1}$, or as
\begin{equation}
\Gamma_{Th}/\Gamma=\sqrt{1+G+G^{2}/2-G^{4}/8}, \label{Eq50}%
\end{equation}
where $G=A/B$. The dependence of the halfwidth of the line on thickness
$T_{A}$ is demonstrated in Fig. 7 showing thickness broadening of the
absorption line. \begin{figure}[ptb]
\resizebox{0.5\textwidth}{!}{\includegraphics{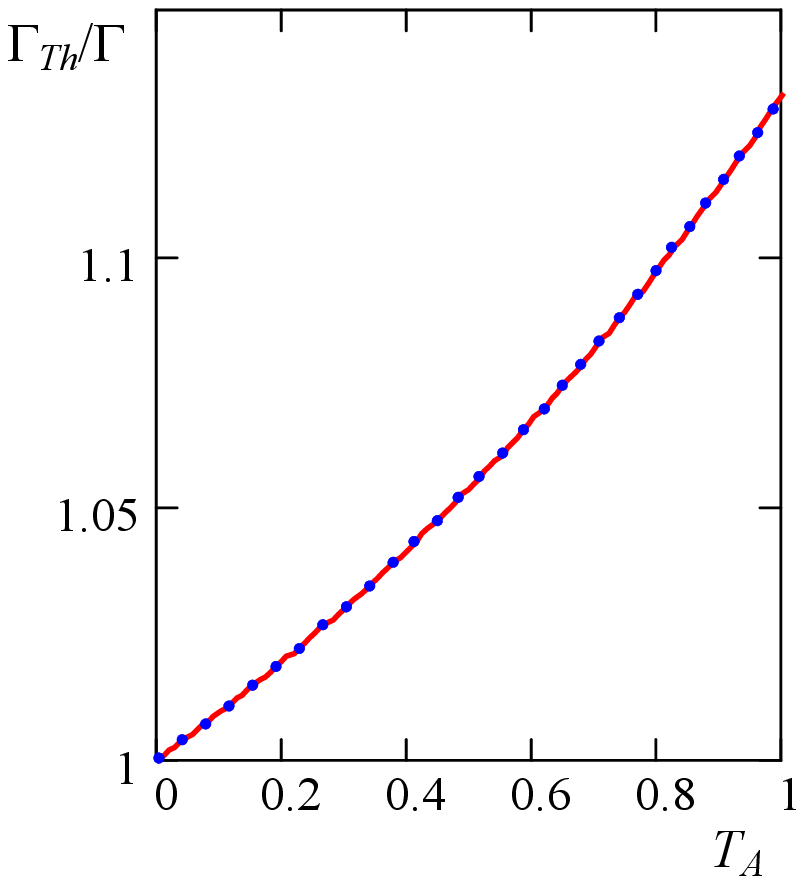}}\caption{Increase of
the line halfwidth of the absorber $\Gamma_{Th}$ with thickness increase.
Solid line (in red) is plotted according to Eq. (\ref{Eq48}) and dotted line
(in blue) corresponds to the approximation Eq. (\ref{Eq49}).}%
\label{fig:7}%
\end{figure}

For a short time interval $(0,T)$ calculation of the integrals in Eq.
(\ref{Eq36}) with the help of approximation (\ref{Eq46}) gives
\begin{equation}
n_{0T}(\Delta)=n_{\infty}(\Delta)\left(  1-e^{-\Gamma T}\right)  +n_{1}%
(\Delta,T)+n_{2}(\Delta,T), \label{Eq51}%
\end{equation}
where%
\begin{equation}
n_{1}(\Delta,T)=\left[  C_{3}(T)\frac{\Gamma\sin\Delta T}{\Delta}%
-C_{4}(T)\frac{1-\cos\Delta T}{\Delta^{2}}-\frac{a_{2}T_{A}^{2}\Gamma T}%
{8}\cos\Delta T\right]  \frac{\Gamma e^{-\Gamma T}}{\Delta^{2}+\Gamma^{2}},
\label{Eq52}%
\end{equation}%
\begin{equation}
C_{3}(T)=C_{1}-\frac{a_{2}T_{A}^{2}}{8}\Gamma T, \label{Eq53}%
\end{equation}%
\begin{equation}
C_{4}(T)=\frac{a_{1}T_{A}}{2}\Delta^{2}+\frac{a_{2}T_{A}^{2}}{8}\Gamma^{2},
\label{Eq54}%
\end{equation}%
\begin{equation}
n_{2}(\Delta,T)=\frac{a_{2}T_{A}^{2}}{4}\left[  1-\cos\Delta T-\left(
1-\frac{\Delta^{2}}{\Gamma^{2}}\right)  \frac{\Gamma\sin\Delta T}{2\Delta
}\right]  \frac{\Gamma^{3}e^{-\Gamma T}}{(\Delta^{2}+\Gamma^{2})^{2}}.
\label{Eq55}%
\end{equation}
Examples of spectra for time intervals $(0,T)$ with different durations of $T$
are shown in Fig. 6 b and c. When $T$ becomes shorter than the lifetime of the
excited state $T_{1}$, the spectrum broadens appreciably.

\subsection{Delayed photon counts}

If we collect photon counts in a delayed time interval $(T,\infty)$ the
spectrum narrows. This spectrum is described by simple equation%
\begin{equation}
n_{T\infty}(\Delta)=n_{\infty}(\Delta)-n_{0T}(\Delta). \label{Eq56}%
\end{equation}
Now, the broad line $n_{0T}(\Delta)$ is subtracted from the narrow line
$n_{\infty}(\Delta)$, which results in the line narrowing effect. Examples of
the line narrowing effect are shown in Fig. 8. \begin{figure}[ptb]
\resizebox{0.5\textwidth}{!}{\includegraphics{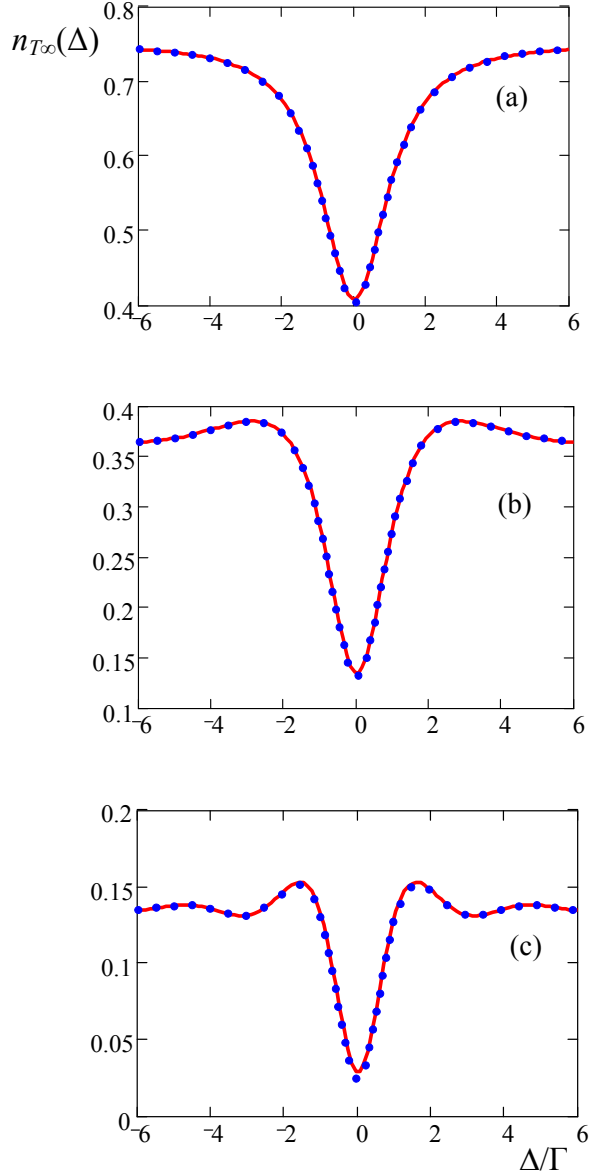}}\caption{Time delyed
spectra. Time $T$ when count collection starts is $0.3T_{1}$ (a), $T_{1}$ (b),
and $2T_{1}$ (c). Solid line (in red) corresponds to exact result. Dotted line
(in blue) is plotted with the help of approximation (\ref{Eq51}). Optical
thickness is $T=1$.}%
\label{fig:8}%
\end{figure}

If we would have a shutter between the source and absorber, which opens at
time $T$, the line narrowing would not appear since photon spectrum, which is
described by Eq. (\ref{Eq11}), coincides with that of the nontruncated photon,
Eq. (\ref{Eq3}), except exponential factor $\exp(-\Gamma_{S}T)$. Therefore, we
conclude that in time delayed spectra the line narrowing appears due to
interference of the coherently scattered field, which takes place at earlier
times, with subsequent portion of the truncated photon wave packet and the
field produced by its scattering. To show that spectrum of photons, emitted at
later time than average, does not narrow, we consider spectra collected in a
short time interval
$\delta\tau=t_{2}-t_{1}$, where $t_{2}>t_{1}$. These spectra are described by
equation%
\begin{equation}
n_{t_{1}t_{2}}(\Delta)=n_{0t_{2}}(\Delta)-n_{0t_{1}}(\Delta).\label{Eq57}%
\end{equation}
Examples of spectra (\ref{Eq57}) are shown in Fig. 9 for $\delta\tau=T_{1}/2$
and $t_{1}$ equal to $T_{1}$ and $3T_{1}/2$. They are broader than steady
state spectra, obtained without delay technique (see Fig. 6a), and have
approximately the same width. \begin{figure}[ptb]
\resizebox{0.5\textwidth}{!}{\includegraphics{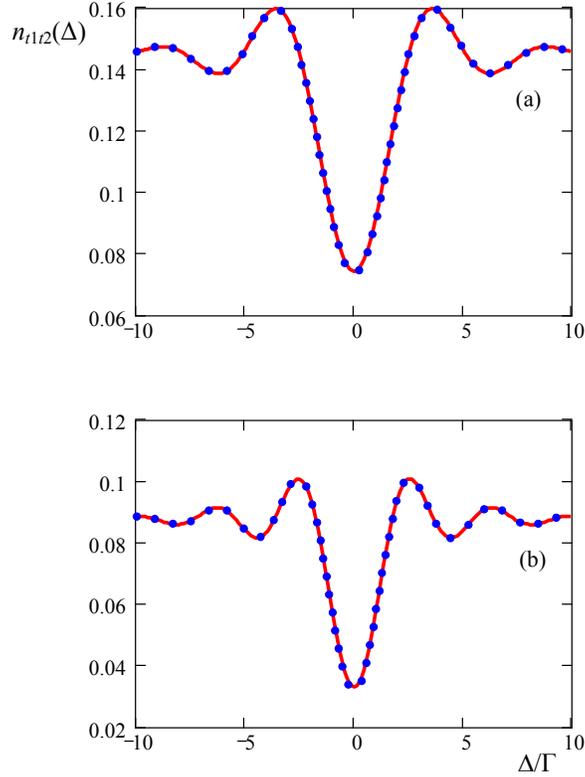}}\caption{Time spectra
collected in a short time interval $(t_{1},t_{2})$ delayed by a long time
$t_{1}$ with respect to formation of excited state nucleus in the source.
Short time interval is $\delta t=t_{2}-t_{1}=T_{1}/2$. Time $t_{1}$ is $T_{1}$
(a) and $3T_{1}/2$ (b). Solid line (in red) corresponds to exact result.
Dotted line (in blue) is plotted with the help of approximation (\ref{Eq51}).
Optical thickness is $T=1$.}%
\label{fig:9}%
\end{figure}

\section{Experiment}

\begin{figure}[ptb]
\resizebox{1\textwidth}{!}{\includegraphics{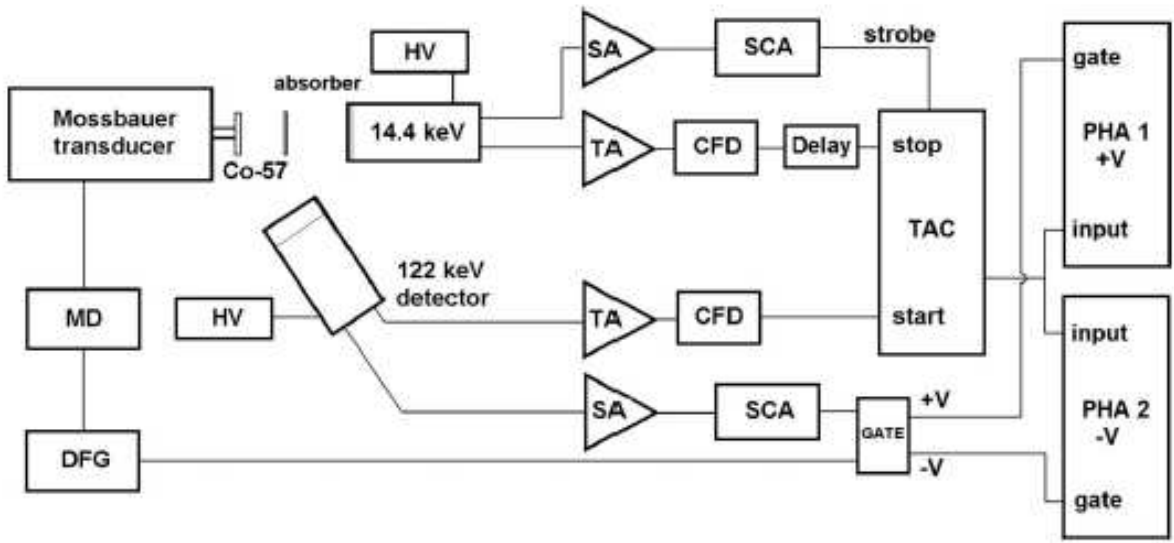}}\caption{Simplified
scheme of the experimental setup. TAC is a time-to-amplitude converter. PHA is
a pulse-height analyzer. TA is a timing amplifier. SA is a spectroscopy
amplifier. SCA is a single-channel analyzer. DFG-MD is the M\"{o}ssbauer
driving unit and function generator. HV is a high-voltage supply.}%
\label{fig:10}%
\end{figure}
Our experimental setup is based on an ordinary delayed coincidence scheme
usually used in measurements of the lifetimes of nuclear states. The schematic
arrangement of the source, absorber, detectors, and electronics is shown in
Fig. 10. The source, $^{57}$Co:Rh, is mounted on the holder of the
M\"{o}ssbauer drive, which is used to Doppler-shift the frequency of the
radiation of the source. The absorber was made of enriched K$_{4}$Fe(CN)$_{6}$
$\cdot$ 3H$_{2}$O powder with effective thickness $T_{A}=13.2$. To calibrate a time
resolution of our setup we measured a time spectrum of the decay of 14.4 keV
state with no absorber. The time resolution of 9.1(5) nsec was obtained by least-squares-fitting the experimental lifetime spectra with the convolution
of the theoretical decay curve and a Gaussian distribution originating from the
time resolution function of the experimental setup (see, for example, Ref.
\cite{Dehaes} for the procedure). The curve measured for a single line source
$^{57}$Co shows the single exponential decay with a mean lifetime $T_{1}=1/\Gamma$
of 140(9) nsec, in good agreement with the mean lifetime and the natural linewidth
data for the 14.4 keV state of $^{57}$Fe.
\begin{figure}[ptb]
\resizebox{0.8\textwidth}{!}{\includegraphics{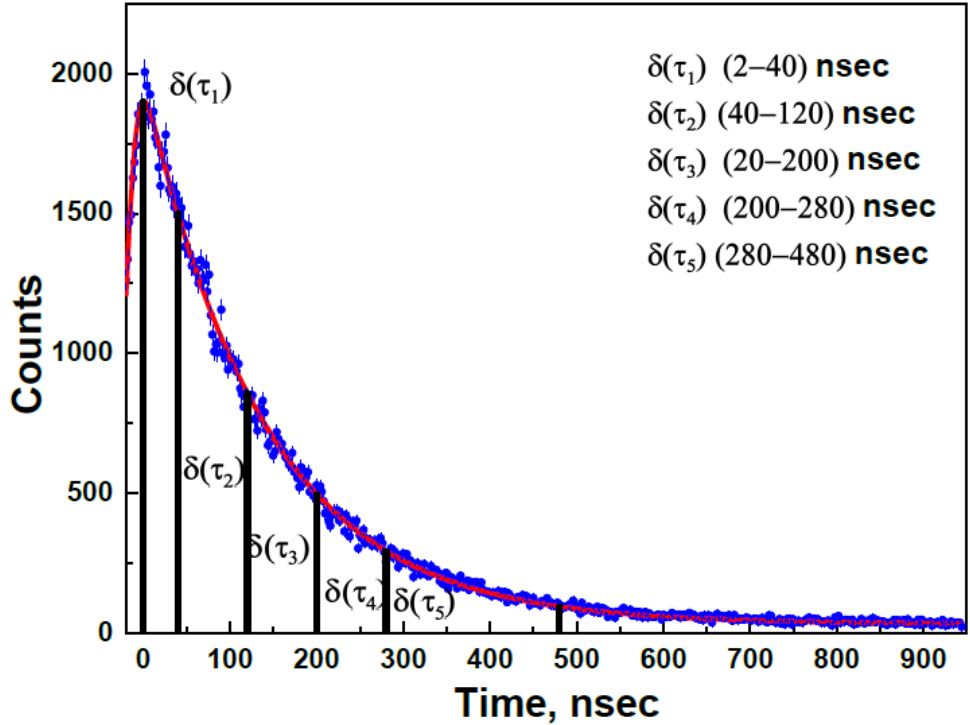}}\caption{Decay curve
of the radiation field from the source along with time windows $\delta \tau_{n}$
where photon counts were collected to plot delayed spectra. Solid line (in read)
is a theoretical fitting by exponent with $T_{1}=141$ ns. Blue dots are
experimental data.}%
\label{fig:11}%
\end{figure}
\begin{figure}[ptbptb]
\resizebox{1\textwidth}{!}{\includegraphics{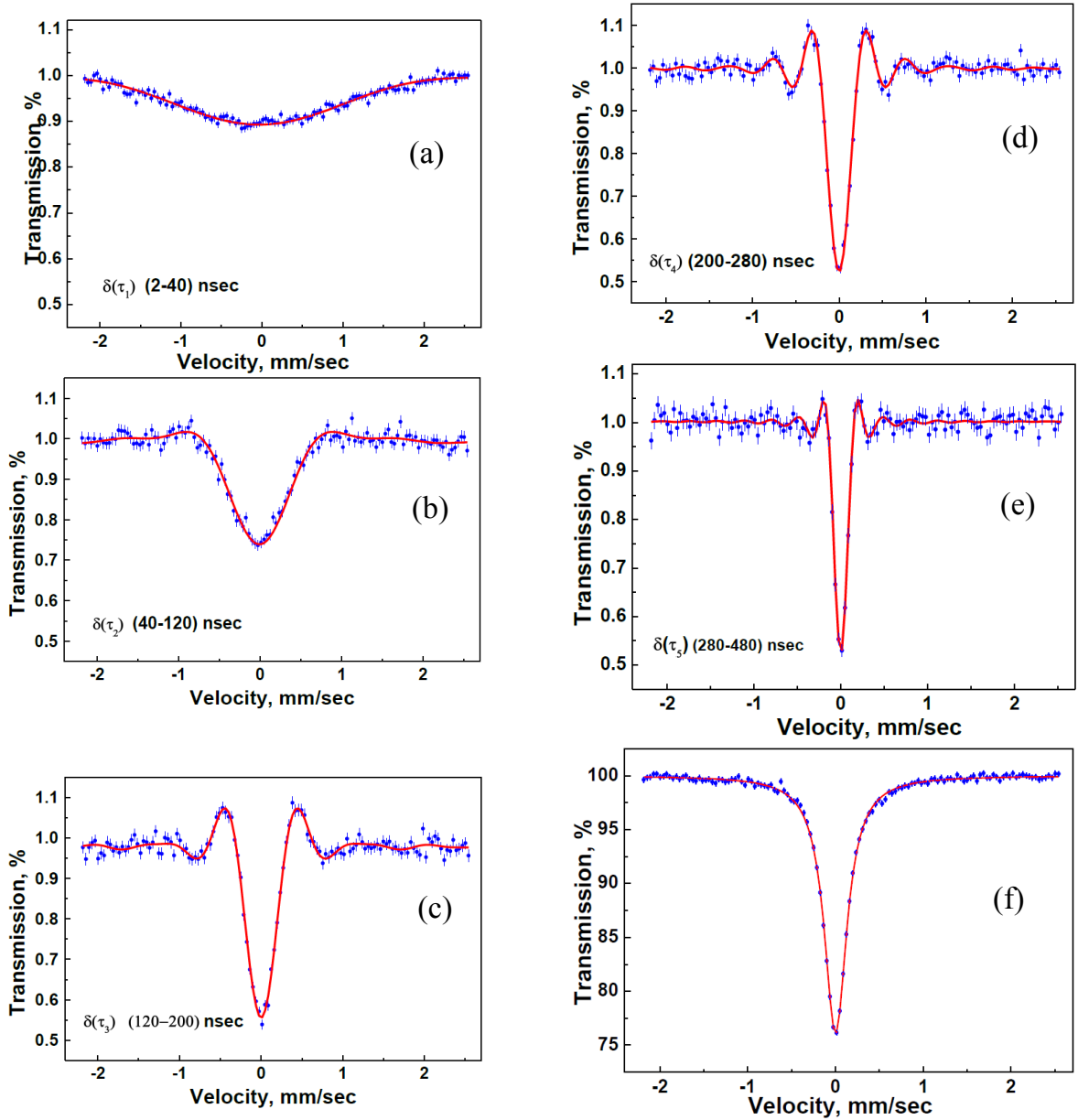}}\caption{Transmiession
spectra obtained with different delay times $t_{1}$ and $t_{2}$ and different
durations of data collection time windows. They are $t_{1}=2$ ns, $t_{2}$ = 40
ns, and $\delta \tau_{1}=38$ ns (a), $t_{1}=40$ ns, $t_{2}=120$ ns, and $\delta
\tau_{2}=80$ ns (b), $t_{1}=120$ ns, $t_{2}=200$ ns, and $\delta \tau_{3}=80$ ns
(c), $t_{1}=200$ ns, $t_{2}=280$ ns, and $\delta \tau_{4}=80$ ns (d), and
$t_{1}=280$ ns, $t_{2}=480$ ns, and $\delta \tau_{5}=200$ ns (e). Conventional
spectrum is shown in (f).}%
\label{fig:12}%
\end{figure}

Decay curve of the radiation field from the source along with time windows
where photon counts were collected to plot delayed spectra are shown in Fig.
11. The background due to accidental coincidences is subtracted from data.
This background is defined from the counting rate at times preceding the fast
front of the incident radiation pulses.

Transmission spectra obtained for different time windows are shown in Fig. 12.
When time window is short, i.e., for $\delta\tau=38$ ns with $t_{1}=2$ ns and
$t_{2}=40$ ns, the line is appreciably broadened (see Fig. 12a). Then, with
increasing $t_{1}$ and lengthening time interval $\delta\tau$, the line
narrows. The narrowest line is obtained for $\delta\tau=200$ ns with
$t_{1}=280$ ns and $t_{2}=480$ ns (see Fig. 12e). It becomes narrower than
usual transmission spectrum obtained without time delayed technique (shown in
Fig. 12f).

Advantage of using this technique we demonstrate here by an example of gamma-resonance measurements with time selection of iron-based nanoparticles on the graphene oxide support. Composite materials, comprising metal nanoparticles on a structural support have gained significant attention in recent years as novel systems for new generation of catalysts, electrode materials in the energy conversion/storage devices, and other similar applications. Due to large surface area, the nano-structured systems possess apparent advantage over the traditional forms of materials. Moreover, iron containing materials are interesting due to their peculiar electro-magnetic and catalytic properties, and low cost/performance ratio.

The graphene oxide (GO) has several apparent advantages as a structural support for growing metal nanoparticles. GO contains oxygen functional groups that facilitate uniform deposition of metals on its surface, and longer exposition, formation and stabilization of metal clusters. Another advantage of GO is its ability to form stable solutions in several solvents by exfoliating to single-atomic-layer sheets. The solution phase provides easy and unimpeded access of reactants to GO surface, opening unlimited venues for liquid phase processing. This property of GO makes it possible to uniformly cover GO surface with nucleation centers of metal nanoclusters made from iron ions present in a bulk salt solution. In our experiment we have used the solution phase reaction between GO and iron (III) nitrate, Fe(NO$_{3}$)$_{3}$.
\begin{figure}[ptbptb]
\resizebox{0.7\textwidth}{!}{\includegraphics{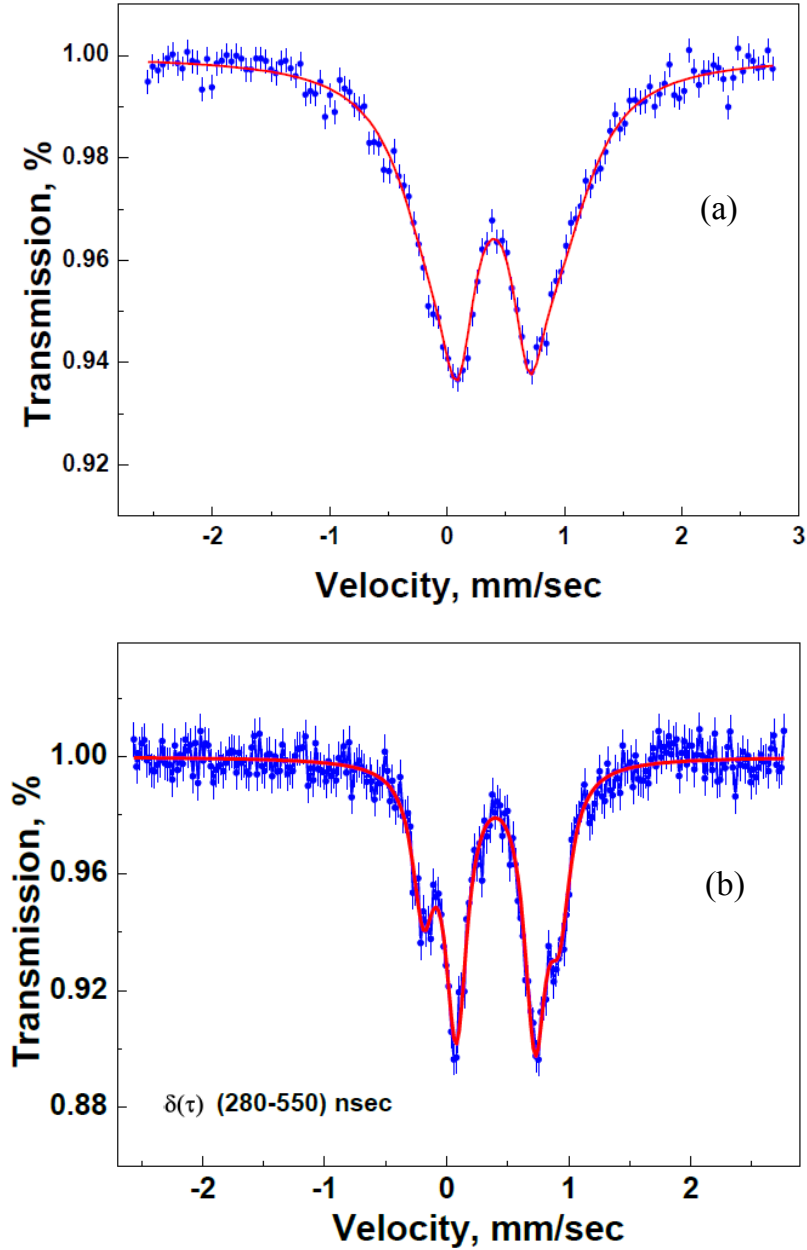}}\caption{Transmiession
spectra of GO containing Fe nanoparticles, which are obtained by conventional
M\"{o}ssbauer spectroscopy (a) and by collecting $\gamma$-photon counts in a time window
$(t_{1},t_{2})$, where $t_{1}=280$ ns and $t_{2}=550$ ns, (b).}%
\label{fig:13}%
\end{figure}

The iron ions uniformly cover GO surface due to the formation of the chemical bonding between Fe(III) cation and GO: the oxygen functional groups of GO serve as ligands, replacing water molecules from the first coordination sphere of the Fe(III) cation. The process can be further complicated by the tendency of Fe(III) to hydrolyze in aqueous solutions. The hydrolysis of Fe(III) ions causes formation on GO surface the hydroxo-complexes of iron, i.e. the clusters where several Fe(III) cations bind to each other via the hydroxide ion bridges (OH$_{-}$).  Undoubtedly, the chemical properties of material will depend on the presence and number of different types of transition metal centers present in it.
	
The M\"{o}ssbauer spectrum of the as-formed Fe-GO sample is shown in Fig. 13(a). It is a quadrupole doublet with slightly asymmetric lines. The doublet hyperfine parameters are IS = 0.36 mm/sec, QS = 0.71 mm/sec, $\Gamma_{1}$ = 0.63 mm/sec and $\Gamma_{2}$ = 0.60 mm/sec, which are the width of the lines in the dublet. The line broadening and their amplitude asymmetry can be explained by the distribution of quadrupole splittings and isomeric shifts, by the Gol'danskii-Karyagin effect due to texture or by the presence of several centers of iron ions. Unfortunately, mathematical processing of the spectrum can not give an obvious preference to one of the above ways of explanation. The M\"{o}ssbauer spectrum of this sample with time selection is shown in Fig. 13 (b). The features associated with the presence of two centers of iron atoms are clearly visible in the spectrum. Probably, in the as-formed Fe-GO there are two types of iron: the Fe(III) ions bonded to the GO functional groups, and the Fe(III) ions bonded to OH$_{-}$ with formation of clusters. Depending on the reaction conditions, the percentage of two types of iron in Fe-GO may vary.

\section{Conclusion}

Delayed photon counts technique is promising in optical and gamma-domain M\"{o}ssbauer spectroscopy. Interference of the radiation fields plays a crucial role in this technique.
In spite of longer time for data collection, this technique is capable to disclose tiny
details of the absorption/transmission spectra if complete information about the energy-level
structure and number of centers contributing to the observable signals are not known in advance.

\section{Acknowledgements}

This work was partially funded by the Russian Foundation for Basic Research
(Grant No. 18-02-00845-a), the Program of Competitive Growth of Kazan Federal
University funded by the Russian Government.

\end{document}